\documentclass[10pt,letterpaper,twocolumn]{article}
\usepackage{ol}

\usepackage{hyperref}
\usepackage{amsmath,amssymb}

\newcommand{\figwidth}{0.45\textwidth}

\begin{document}

\twocolumn%
[

\newcommand{\ethaffil}{Laboratory of Physical Chemistry, Swiss Federal Institute of Technology (ETH), 8093 Z\"urich, Switzerland}
\newcommand{\mariaaffil}{Institute of Electronic Structure and Laser (IESL), Foundation for
Research and Technology Hellas (FORTH), P.O. Box 1527, 71110
Heraklion, Crete, Greece, and Dept. of Materials Science and Technology,
Univ. of Crete, Greece.}

\title{Spontaneous emission rates of  dipoles in   photonic crystal membranes }

\author{A. Femius {Koenderink}}
\affiliation{\ethaffil}
\author{Maria Kafesaki}
\affiliation{\mariaaffil}
\author{Costas M. Soukoulis}
\affiliation{Ames Laboratory, Iowa State University, Ames, Iowa 50011, and \mariaaffil}
\author{Vahid Sandoghdar}
\affiliation{\ethaffil}

\date{Prepared December 6, 2005.}

\begin{abstract}
We show theoretically that   finite two-dimensional (2D)  photonic
crystals in thin semiconductor membranes strongly modify the
spontaneous emission rate of   embedded dipole emitters.
Three-dimensional Finite-Difference Time-Domain calculations show
over 7 times inhibition and   15 times enhancement of the emission
rate compared to  the vacuum emission rate for judiciously
oriented and positioned dipoles. The vertical index confinement in
membranes strongly enhances modifications of the emission rate as
compared to vertically unconfined 2D photonic crystals. The
emission rate modifications inside the membrane mimic the local
electric field mode density in a simple 2D model. The inhibition
of emission saturates exponentially as  the crystal size around
the source is increased, with a $1/e$ length that is inversely
proportional to the bandwidth of the emission gap. We obtain
 inhibition of emission  only close to the slab center. However, enhancement
of emission persists even outside the membrane, with a distance
dependence which dependence can be understood by analyzing the
contributions to the spontaneous emission rate of the different
vertically guided modes   of the membrane.
 Finally we show that the emission changes can even be observed in experiments with
 ensembles of randomly oriented dipoles, despite the contribution
of dipoles for which no gap exists.
\end{abstract}

]

\section{Introduction}

 The last decade has seen a remarkable increase in the experimental efforts
 to control spontaneous emission dynamics of dipole transitions by
 tailoring the dielectric surroundings of the source. While
studies in elementary geometries, such as  planar
 interfaces\cite{drexhage70,khosravi91,snoeks95,amosandbarnes,buchler05,qbook}
 or   spherical  particles\cite{chew,klimov,hermannssphereref,schniepp}, remain of current interest,
 particular impetus in spontaneous emission control  derives  from   solid state cavity quantum
 electrodynamics (CQED)\cite{qbook,vahalareview}, and the increasing capability to
 shape semiconductors  on the nanometer scale. Of
 particular interest in this respect are  photonic band gap
 materials\cite{soukoulis01}.
 As first proposed by  Bykov\cite{bykov72}
and Yablonovitch\cite{yablonovitch87}, three-dimensionally
periodic arrangements of dielectric material on wavelength-sized
length scales  allow to create a  medium in which spontaneous
emission processes can not only be enhanced  but also completely
inhibited.  These  photonic crystals offer many exciting prospects
in fundamental physics, ranging from CQED, control of black-body
radiation\cite{blackbody}  and the slowing and storage of
light~\cite{slowing}, as well as device
opportunities~\cite{soukoulis01} for producing and processing of
optical signals on sub-micron length scales.

Although efforts in realizing spontaneous emission control in
three-dimensional (3D)  photonic band gap crystals have been
partly successful\cite{koenderink02,lodahl04,ogawa04}, the
advantages of easier fabrication and characterization has
motivated many groups to work on  lower-dimensional, i.e.,
two-dimensional  (2D)   crystal structures. Photonic crystals in
thin semiconductor membranes have recently shown particular
promise  to realize   ultrasmall cavities with very high quality
factors, suited for quantum optics in the strong coupling
limit\cite{nodacav,srinivasan,yoshie}.  In addition, the first
experimental evidence\cite{finley,vuckovicemits,nodainhibit} for
large emission inhibition of quantum dots and quantum wells in
photonic crystal membranes has recently been reported. Although
all these results suggest that the effect of the gap in the
in-plane band-structure is greatly enhanced by the vertical index
contrast, a systematic study of the potential for spontaneous
emission rate control offered by membrane photonic crystals is
lacking.

In this Paper we present a  theoretical study of the spontaneous
emission rate modifications   in   membrane photonic crystals that
consist of only a  finite number of unit cells. Such an analysis
is particularly challenging for structures that are
 limited in size because approaches
that are based on the commonly used plane-wave electromagnetic
eigenfunction
calculations\cite{ho90,suzuki95,busch98,busch00,li00,wanggrouptheory}
cannot be used. We have therefore used a method similar to that
recently proposed by Hwang et al.\cite{hwang99}, Xu et
al.\cite{xu00,lee00} and Hermann and Hess\cite{hermann}, based on
the three-dimensional  finite difference time-domain method
(FDTD)\cite{taflove}.

The paper is organized as follows: in Section~\ref{method} we
summarize our theoretical approach and discuss the numerical
characteristics of the method. In Section~\ref{comparison} we
discuss the spontaneous emission lifetime modification   for a
dipole in the center of a photonic crystal membrane, and compare
our result to both 2D and 3D plane wave calculations of the local
density of states. In Section~\ref{polarization} we focus on the
dependence of the emission lifetime on the  orientation of the
transition dipole moment and the position  of the emitter in the
central plane of the membrane. In section~\ref{gapwidth} we
consider the role of the crystal size in determining the
spontaneous emission rate, and propose a relation between the
width of the emission gap and the number of crystal rows that need
to surround the emitter to obtain maximum inhibition. In
section~\ref{sepdep}, we focus on the effect of membrane thickness
and of the height of the dipole above the membrane center on the
emission lifetime. We show that a simple model for the emission
rate in homogeneous dielectric slabs explains many of the observed
features. Finally, in Section~\ref{ensemble} we discuss the
prospects for experiments   involving localized ensembles of
emitters with randomly distributed dipole moments. Conclusions are
presented in Section~\ref{conclusions}.

\section{Method\label{method}}

The quantum analysis of spontaneous emission based on Fermi's
Golden Rule asserts that the spontaneous emission rate of an
emitting dipole varies with its position $\mathbf{r}$ and
orientation $\mathbf{\hat{d}}$, depending on the availability of
photon modes at the source position and frequency $\omega$. The
central quantity that takes this into account is the local
radiative density of states\cite{sprik96} (LRDOS)
$\rho(\mathbf{r},\hat{\mathbf{d}},\omega)$. The LRDOS is a
classical electromagnetic quantity that appears in the classical
analysis of radiating antennas, where it describes the power
needed to drive a point-like electric dipole oscillating at fixed
current\cite{hwang99,xu00,lee00,hermann}. This power  is due to
the work  $W\propto \mathbf{j}\cdot\mathbf{E}$ that the field
$\mathbf{E}$  radiated by the source $\mathbf{j}(\mathbf{r},t)$
does on the source itself. The rate of spontaneous emission
$\Gamma$ in a dielectric structure, normalized to the free-space
rate $\Gamma_{\rm vac}$, therefore can be found by comparing the
work $W$ done on the source in the dielectric structure to the
work $W_{\rm vac}$ done on the same current source in vacuum.

\begin{figure}[tb]
\centerline{\includegraphics[width=\figwidth]{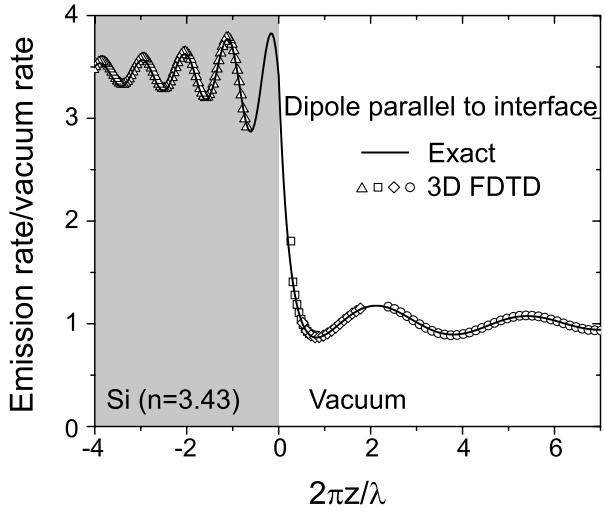}}%
\caption{Symbols: spontaneous emission rate normalized to emission
rate in vacuum as a function of normalized frequency  for a dipole
in front of, and oriented parallel to  a planar dielectric
interface separating silicon ($\epsilon=11.76$) and vacuum
calculated using the 3D FDTD   method. The normalized frequency
corresponds to the separation $z$ between source and interface,
normalized to the vacuum emission wavelength $\lambda$.  For the
dipole in vacuum (positive $z$), results are shown for three
separations. In  units of the grid discretization $\Delta$, these
are  $z=25\Delta$ (circles), $5\Delta$ (diamonds) and $2\Delta$
 (squares). The same temporal excitation pulses were used
 for $z=25\Delta$ and $5\Delta$  to span two different normalized frequency ranges $2\pi z/\lambda$.
  For the dipole  in silicon (negative $z$) the results shown were
  obtained with a single excitation pulse and a single  dipole position at  $z=-25\Delta$  from the interface.
   Excellent agreement with the exact local radiative density of states modification (solid
 line) is obtained. The total simulation volume was $300\times 300\times 120~\Delta^3$ in size.
 \label{fig:slabldos}}
\end{figure}

We use the three-dimensional finite different time domain method
(FDTD) method for our calculations throughout this work. An
important advantage of the finite difference time domain
method\cite{xu00,lee00,hermann,taflove} over conventional
methods\cite{khosravi91,busch98,busch00} to calculate the LRDOS is
that the field $\mathbf{E}$ of a point source, and hence the
LRDOS, is readily calculated without solving for the
electromagnetic eigenmode basis of the dielectric structure.
Furthermore, the FDTD method can even be used to calculate the
LRDOS   for lossy systems in which eigenmode expansions are
invalid. In this paper we follow the method of Hermann and
Hess~\cite{hermann}, in which the LRDOS is simulated for a wide
frequency range at once via broadband temporal excitation of a
dipole point source. For a pulsed source   with frequency
 spectrum $\mathbf{j}(\mathbf{r}_0,\omega)$, the
time-trace   of the electric field $\mathbf{E}(\mathbf{r}_0,t)$ at
the position $\mathbf{r}_0$ of the source can be Fourier
transformed to determine the  LRDOS in terms of its vacuum value
$\rho_{\rm vac}(\omega)$:
\begin{eqnarray}
\rho(\mathbf{r}_0,\hat{\mathbf{d}},\omega)&=&\rho_{\rm
vac}(\omega)\frac{W(\mathbf{r}_0,\hat{\mathbf{d}})}{W_{\rm
vac}}\nonumber\\
&=&\rho_{\rm vac}(\omega)
\frac{\mathbf{j}(\mathbf{r}_0,\omega)\cdot\mathbf{E}(\mathbf{r}_0,\omega)}{\mathbf{j}(\mathbf{r}_0,\omega)\cdot\mathbf{E}_{\rm
vac} (\mathbf{r}_0,\omega)} \label{eq:FDTDLDOS}\end{eqnarray} We
have evaluated the validity and pitfalls of this numerical method
by comparing the simulated spontaneous emission rate enhancement
with exact analytical results for a dipole in a dielectric
sphere\cite{chew,klimov,hermannssphereref,hermann}, and for a
dipole near a planar interface between two
dielectrics\cite{khosravi91,snoeks95,amosandbarnes}.
Figure~\ref{fig:slabldos} shows that accurate results for the
emission rate modification are obtained  by the FDTD method for a
representative test case   where  a dipole is placed near a planar
interface between air and silicon. By examining the LRDOS obtained
with pulses at different center frequencies,  we have established
that results consistent to within $\sim 1\%$ can be obtained in
the frequency ranges where the vacuum emission power of the dipole
current pulse exceeds $\sim 5\%$ of its maximum value. Throughout
this paper we use temporal excitation pulses with a Gaussian
envelope and bandwidths around 50\% of the central frequency.

\begin{figure}[tb]
\centerline{\includegraphics[width=\figwidth]{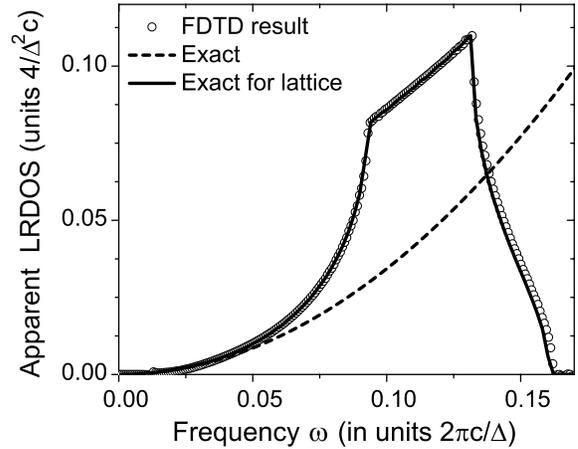}}
\caption{Symbols: FDTD approximation to the  LRDOS of a
\emph{homogeneous} medium ($\epsilon=11.76$) obtained using the
FDTD method and Eq.~(\ref{eq:FDTDLDOS}), taking a very broadband
temporal excitation of the source dipole. The dashed parabola
indicates the exact LRDOS. Agreement   to within $5\%$ are only
obtained for frequencies $\omega <0.03 (2\pi c)/\Delta$, for which
the wavelength in the highest index medium remains longer than
$10$ times the discretisation grid spacing $\Delta$. Deviations
are due to the unphysical dispersion relation on the grid
(Eq.~(\ref{eq:FDTDdispersion})) and the wave vector cut-off of the
grid at $k=(\pi/\Delta,\pi/\Delta,\pi/\Delta)$, as evident from
the excellent agreement of the FDTD result with the exact
\emph{grid} density of states per volume obtained by integrating
Eq.~(\ref{eq:FDTDdispersion}). \label{fig:griddos}}
\end{figure}

Agreement of the simulated LRDOS with the exact LRDOS not only
depends on sufficient spectral overlap of the input current with
the frequency range of interest, but also  on the spatial
discretisation. Apart from the obvious requirement that the
discretisation must be sufficiently fine to resolve features of
the dielectric structure, the accuracy is affected by the
dispersive properties of the cubic discretisation grid itself. To
illustrate this point, Fig.~\ref{fig:griddos} shows
 the apparent LRDOS of a homogeneous dielectric of index
$n=3.4$ over a very wide frequency range, obtained by the
procedure described in Eq.~(\ref{eq:FDTDLDOS}). The exact LRDOS of
the homogeneous dielectric (parabola in Fig.~\ref{fig:griddos})
equals $n$ times the vacuum LRDOS, independent of frequency. In
contrast, Fig.~\ref{fig:griddos} shows that  large discrepancies
occur for large frequencies. The deviation exceeds 5\%   for
frequencies at which the wavelength in the dielectric is shorter
than $10\Delta$, where $\Delta$ is the spacing between adjacent
discretisation grid points. The source of the discrepancy is the
well-known anisotropic dispersion of the cubic   grid as well as
the cut-off of the grid for wave vectors in excess of
$(\pi/\Delta,\pi/\Delta,\pi/\Delta)$. We have calculated the
density of states of the discretisation grid (filled with vacuum
and with medium of index $n$) from its dispersion
relation,\cite{taflove}
\begin{eqnarray}
\sin\left(\frac{\omega\Delta t}{2}\right)^2&=&\frac{c\Delta
t}{n\Delta x}\Big[\sin\left(\frac{k_x\Delta }{2}\right)^2+ \nonumber\\
 & &  \sin\left(\frac{k_y\Delta }{2}\right)^2
 +\sin\left(\frac{k_z\Delta }{2}\right)^2\Big],\label{eq:FDTDdispersion}\\
\nonumber\end{eqnarray}%
 where $\Delta t$ is the
magnitude of the time step, $\Delta$ is the grid spacing, and $c$
is the speed of light in vacuum. The excellent agreement of the
grid density of states with the apparent LRDOS calculated by the
FDTD method for a dipole in a homogeneous dielectric in
Fig.~\ref{fig:griddos}, confirms the role of the density of states
of the numerical discretisation grid in spontaneous emission rate
calculations. We conclude that an accuracy better than $\sim 5\%$
for  spontaneous emission rates calculated by the FDTD method,
requires discretisation grids with over  10 grid points per
wavelength at the frequency of interest in the highest index
medium.

Finally we note that accurate results require that the Fourier
transforms of $\mathbf{j}(\mathbf{r}_0,t)$ and
$\mathbf{E}(\mathbf{r}_0,t)$ be  calculated without zero-padding
of the time series, as done by many FFT routines that use arrays
of length equal to a power of $2$. To find the power spectrum
radiated by the dipole, the complex Fourier transforms ${\cal F}
$of the real time series of $\mathbf{j}$ and $\mathbf{E}$ are
combined according to
\begin{equation}
\mathbf{j}(\mathbf{r}_0,\omega)\cdot\mathbf{E}(\mathbf{r}_0,\omega)={\rm
Re}[{\cal F}\mathbf{j} ]\cdot{\rm Re}[{\cal F}\mathbf{E} ]+{\rm
Im}[{\cal F}\mathbf{j} ]\cdot{\rm Im}[{\cal F}\mathbf{E}].
\label{eq:noise}\end{equation}
 The two terms are nearly equal in magnitude but opposite in
sign. The desired sum is approximately $10^4$ times smaller than
either term. Even if time-stepping is continued till the fields
have decayed by several orders of magnitude,  zero padding
introduces noise that is small compared to either term in
Eq.~(\ref{eq:noise}), but sizeable compared to the signal that
remains after summing them. Furthermore,  strongly resonant
features in  LRDOS spectra need to be handled with care. Time
series of field and current are typically limited to total time
spans of only $10^2$ to $10^3$ optical cycles, as the 3D FDTD
method  is very computationally intensive. For resonances with
quality factors in excess of $\sim 10^2$, time series are
truncated before fields have fully decayed, giving rise to
artifacts in the Fourier transform that can fully dominate the
simulated LRDOS spectra. To remove such artifacts, we have used a
seventh order Daubechies apodization filter~\cite{daubechies}. We
have checked the validity of this approach for FDTD simulations of
a single-mode high-Q photonic crystal cavity~\cite{nanoswitch},
which has a much narrower resonance in the LRDOS than appears in
any of the results presented below. For such a single mode system,
the time series of the field can be extrapolated to obtain the
true LRDOS from an untruncated time series.

\section{Comparison with 2D and 3D LRDOS for vertically infinite crystals\label{comparison}}

We consider  spontaneous emission rate modifications in
two-dimensional photonic crystals within semiconductor membranes,
as studied for the dipole-position averaged case by Lee and
Yariv~\cite{lee00}. We focus on finite structures consisting of
membranes with dielectric constant $\epsilon=11.76$ and thickness
$250$~nm perforated with a hexagonal lattice of holes (lattice
spacing $a=420$~nm) of radius $0.3a$. Such membranes have a wide
band gap for TE guided modes, i.e., modes with electric field in
the central plane of the slab pointing along the membrane, but no
gap for TM modes. The geometrical parameters are typical for the
membrane structures in which high-quality factor cavities for
strong coupling CQED were recently
realized\cite{nodacav,srinivasan,yoshie}  and  for which first
evidence of inhibition of spontaneous emission was recently
presented\cite{finley,vuckovicemits}.
\begin{figure}[tb]
\centerline{\includegraphics[width=\figwidth]{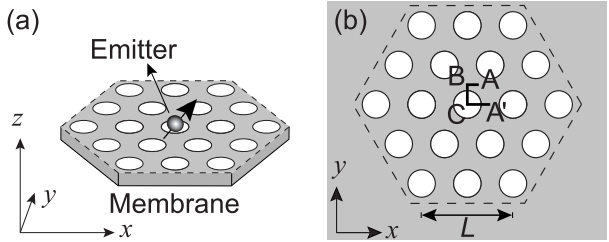}}\caption{(a)
Schematics of the geometry for calculating the spontaneous
emission rate modification in membranes. We consider   dipoles
near the center of 2D finite photonic crystals, consisting of a
hexagonal array of air holes (lattice spacing $a$, radius $r$) in
a semiconductor membrane (thickness $d$). The position and
orientation of the emission dipole are specified along cartesian
axes, where  $z$ specifies the height above the center of the
membrane, and $x$ and $y$ are  in the plane of the membrane. (b)
(top view) The finite crystallites are hexagonally truncated, such
that the central air hole is surrounded by the same number of
holes (indicated by the crystallite radius $L$ ($L=2a$ in the
sketch)) in all directions. Outside the truncation boundary
(dotted hexagon), the membrane was either terminated by
semiconductor or by air. For Figure~\ref{fig:contourLDOSmembrane},
the dipole position was scanned over the trajectory from point $A$
via $B$, $C$ to $A'$. \label{fig:structure}}
\end{figure}
 We
simulate finite crystallites of hexagonal shape and various sizes
ranging from 1 to 12 lattice spacings in radius, as counted from
the central hole (see Fig.~\ref{fig:structure}). The crystals were
either terminated by semiconductor or by air extending into the
Liao absorbing boundary conditions. We have used 14 or 20 grid
points per lattice constant, with a doubled grid density in the
vertical direction. To reduce staircasing errors due to the
discretisation of the structure, we   used volume-averaging of the
dielectric constant in each grid cell\cite{hermannpssa}. In the
simulations, membranes were surrounded above and below by up to
1$~\mu$m of air.
\begin{figure}[tb]
\centerline{\includegraphics[width=\figwidth]{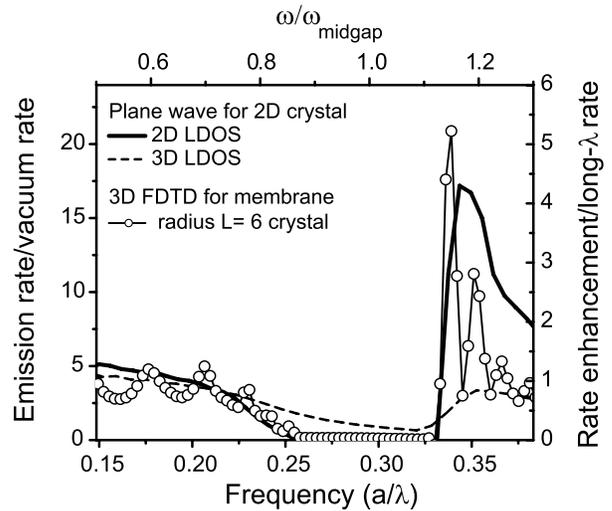}}\caption{Connected
open symbols: modification of the radiative emission rate compared
to vacuum (refer to left axis) versus normalized frequency
$a/\lambda$ (refer to bottom axis). This result is calculated with
the 3D FDTD method for a dipole in the central air hole of a
 crystal of radius $L=6a$ with $r/a=0.3$, $a=420$ in a semiconductor membrane
 ($\epsilon=11.76$, $d=250$ (truncated by semiconductor)), oriented along the $x$-axis in Fig.~\ref{fig:structure}, in the plane of the membrane. In the
 apparent gap, emission is   reduced to about 0.14 times the vacuum emission
 rate.  Refer to the top- and right-hand axis for frequency
 normalized to the midgap frequency, and emission rate enhancement
 normalized to the emission rate modification at the same position
 for frequencies far below the gap.
Solid line (refer to top- and right-hand axis): modification of
the 2D LRDOS in an infinite crystal (also $r/a=0.3$,
$\epsilon=11.76$) in the center of an air hole, according to the
plane wave method.  Dashed line: modification of the 3D LRDOS in
an infinite 2D crystal (again $r/a=0.3$, $\epsilon=11.76$) that
extends infinitely in the vertical direction, calculated with the
plane wave method. The LRDOS modification in the center of the
membrane is much stronger than the weak LRDOS modification in the
vertically unconfined crystal. \label{fig:centralinhib}}
\end{figure}

 First, we consider the spontaneous emission
lifetime for a dipole that is located in the middle of the central
air hole of a structure of $L=6$ lattice spacings across.
Figure~\ref{fig:centralinhib} shows the LRDOS normalized to the
vacuum LRDOS for an in-plane oriented dipole that points along the
$x$-axis (see Fig.~\ref{fig:structure}), i.e., from air hole to
air hole.  For dipole orientations in the plane of the slab, the
spontaneous emission rate is strongly modulated as a function of
normalized frequency $a/\lambda$ (where $\lambda$ is the vacuum
wavelength). We find a frequency range from $a/\lambda=0.25$ to
$a/\lambda=0.33$ of
 inhibition of the emission rate that overlaps with the in-plane photonic band
gap for the TE modes  guided by the membrane, as was also observed
in calculations by Lee, Xu and Yariv~\cite{lee00}. An inhibition
of the LRDOS by a factor $\gtrsim 7$ compared to vacuum is
obtained. This  value translates to over $30$ times inhibition as
compared to the emission modification at frequencies far below the
band gap at the same position. Above and below the range for which
the emission is strongly reduced, which we will call
\emph{emission gap}, the emission rate is strongly modulated,
featuring enhancements of the emission rate by factors up to 20
times compared to vacuum, or over $5$ times  compared to the LRDOS
in the long wavelength limit at the same position. As will be
discussed below, the fringes outside the emission gap are due to
Fabry-P\'erot oscillations caused by the finite lateral size of
the crystal.

It is perhaps surprising  that significant control over
spontaneous emission dynamics is possible for selected dipole
positions and orientations, despite the fact that no full band gap
exists. It appears that the large vertical index contrast greatly
enhances the effect of the in-plane periodicity on spontaneous
emission. To confirm this notion, we have used the H-field
inverted matrix plane-wave expansion method to calculate the local
radiative density of states in infinite 2D photonic crystals that
are not confined vertically\cite{ho90,busch98,busch00}. In
addition to the 3D LRDOS, wee have also calculated  the 2D LRDOS,
which only counts modes that propagate in the plane of
periodicity.  For the two (three)-dimensional case, we used 8100
(84000) non-equivalent k-points\cite{gilat72,monkhorst76}, which
were distributed over half of the full Brillouin zone to avoid
erroneous results\cite{busch98} due to the reduced symmetry of the
polarization-resolved LRDOS\cite{wanggrouptheory}.  We have scaled
both the frequency axis and vertical axis to optimally overlap
 the resulting 2D and 3D LRDOS (normalized by the 2D resp. 3D vacuum LRDOS)
 for an $x$-oriented dipole in the center of an air-hole    with
 the LRDOS in the membrane (cf. Fig.~\ref{fig:centralinhib}).
 Naturally, the
2D (local) density of states of the 2D crystal shows a full gap
for dipole orientations in the plane of periodicity, corresponding
to the TE band gap in the 2D band structure. The gap is
accompanied by characteristic emission enhancement at the gap
edges due to low group velocity modes\cite{busch98}. For a dipole
at the center of an air hole, this enhancement occurs at the high
frequency edge of the gap, as is also apparent in the membrane. In
stark contrast, the 3D LRDOS is only weakly modulated for
vertically infinite crystals, due to the contributions of modes
with wave vector components along the air
cylinders\cite{asatryan01,fussell03}. Although the 2D band gap
causes a shallow minimum in  the 3D LRDOS for a dipole in the
center of an air-hole, the bandwidth of emission inhibition is
much narrower than the bandwidth of the 2D band gap, and no sign
of emission enhancement at the band gap edges
remains\cite{busch98,fussell03}. The comparison of plane wave
calculations with the FDTD result
 shows that  the vertical confinement offered by membranes
prevents the washing out of the characteristic features of the 2D
LRDOS that occurs for vertically unconfined crystals: Firstly  the
emission gap in the membrane  referenced to the long wavelength
LRDOS at the same position in the crystal  amounts to an
inhibition by a factor around 30, which is over five times
stronger than the inhibition obtained if the structure is not
truncated in the vertical direction. Secondly, the relative
inhibition gap bandwidth  of 24\% offered by the membrane
structure is close to the bandwidth of 25\% in the 2D
band-structure. Thirdly, we note that the emission rate
enhancement at the blue edge of the gap is retrieved in the
membrane structure. The enhancement of $\sim 6$ times compared to
the emission rate in the long wavelength limit is in good
agreement with the enhancement observed in the 2D LRDOS. This is
especially surprising, when considering the finite size of the
crystal, and the effect of its truncation   on the exact spectral
position and quality factor of the Fabry-P\'erot oscillations.

\section{Dependence on polarization and lateral position\label{polarization}}

\begin{figure}[tb]
\centerline{\includegraphics[width=\figwidth]{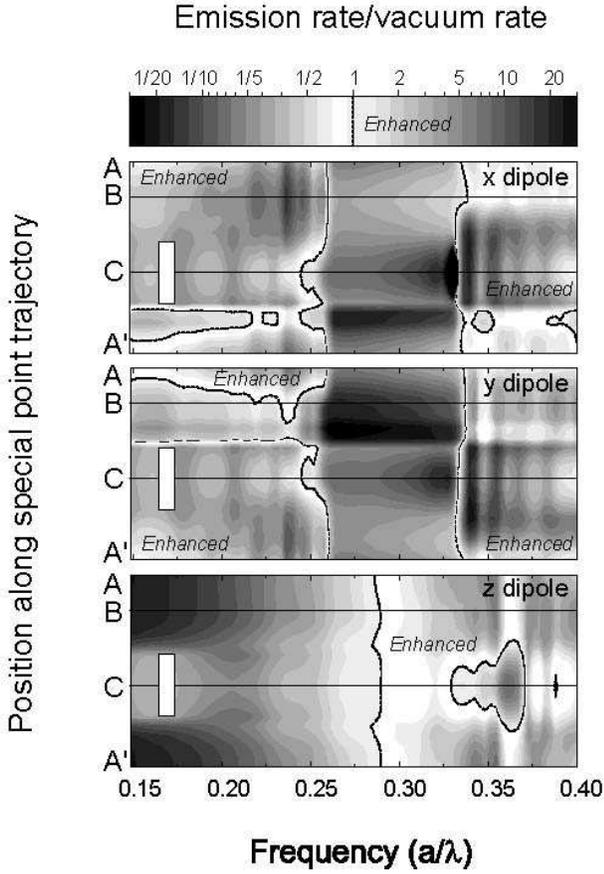}}\caption{Contour
plot of the LRDOS  normalized to the vacuum LRDOS   calculated by
3D FDTD as a function of frequency and as a function of position
along the trajectory in Fig.~\ref{fig:structure} for dipoles in a
finite membrane crystallite of radius $L=6a$. Panels (a--c) refer
to the three orthogonal dipole orientations along $x,y$ resp. $z$.
Vertically, the dipoles are placed in the middle of the membrane
($z=0$). In-plane dipole orientations show a large emission gap
for $0.25<\omega<0.33$ for all positions, while no gap occurs for
the vertically oriented dipole. Enhancement of the emission rate
for horizontally oriented dipoles occurs on the high frequency
edge when the dipole is located in air (positions indicated by the
vertical white bars in all panels) and on the low frequency edge
when the dipole is located in dielectric. The grayscale (indicated
on top) is logarithmic both for inhibition and enhancement
(regions separated by contour line at unity, and labelled as
`Enhanced' as appropriate). \label{fig:contourLDOSmembrane}}
\end{figure}
 To obtain a more complete overview of the
spontaneous emission control offered by membranes, we consider the
LRDOS modification  for different positions in the central unit
cell of a crystallite that is  again $6a$ in radius, terminated on
the sides by semiconductor material.
Figure~\ref{fig:contourLDOSmembrane} shows contour plots of the
LRDOS normalized to the vacuum LRDOS as a function of frequency in
the range of the TE-bandgap, for dipoles that are vertically
centered in the membrane and positioned along the path indicated
in Fig.~\ref{fig:structure}(b). This trajectory spans the boundary
of the irreducible part of the real space 2D unit cell and ensures
that the cartesian dipole orientations in
Fig.~\ref{fig:contourLDOSmembrane} are always
 parallel or perpendicular to the trajectory.
Figure~\ref{fig:contourLDOSmembrane} confirms   the occurrence of
a deep gap in the spontaneous emission rate  for dipole
orientations in the plane of the membrane    for all positions
close to the center of the crystal.  Moreover we find that there
is no gap in the emission rate for dipoles oriented perpendicular
to the membrane, but   only a weak dependence on frequency, as
expected from the absence of a 2D band gap for the TM guided modes
of the membrane.

\begin{figure}[tb]
\centerline{\includegraphics[width=\figwidth]{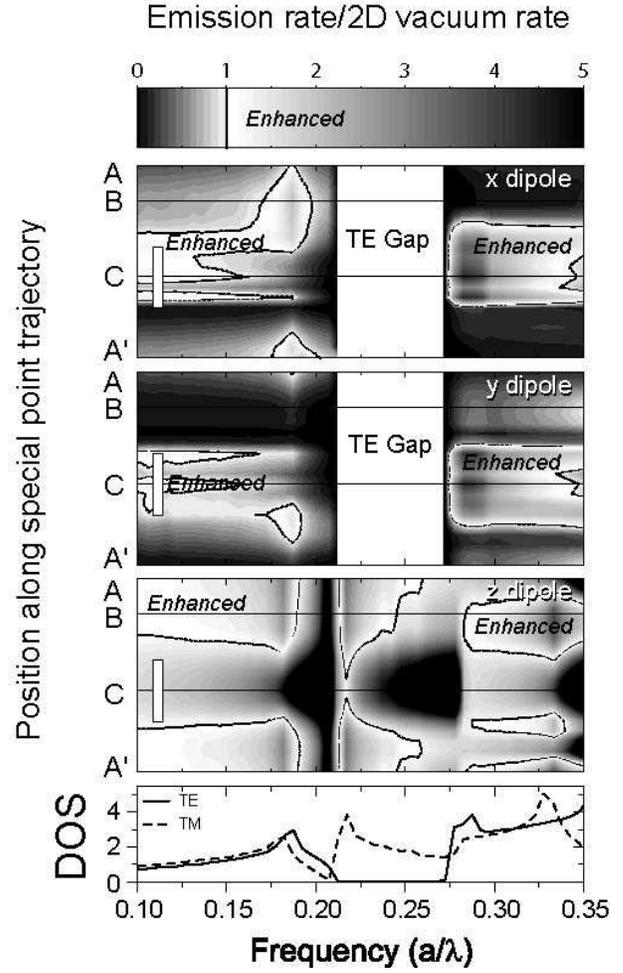}}\caption{Top
three panels: Contour plots of the 2D LRDOS normalized to the
vacuum 2D LRDOS according to the plane wave method  for a dipole
in a 2D photonic crystal (air holes $r/a=0.30$ in a matrix of
$\epsilon=11.76$) as a function of emission frequency (horizontal
axis) and position of the dipole along the trajectory indicated in
Fig.~\ref{fig:structure} for three perpendicular dipole
orientations as indicated. White bars indicate the extent of the
trajectory inside air. For in plane oriented dipoles (top two
panels), the 2D LRDOS shows a full gap (labelled as TE gap). The
bottom panel shows the 2D TE and TM total density of states per
area in units $a^{-1}c^{-1}$. \label{fig:2dpw}}
\end{figure}

 A detailed inspection of  Fig.~\ref{fig:contourLDOSmembrane}
shows furthermore that strong emission enhancement at the blue
edge of the emission gap is only predicted to occur for dipoles
positioned inside the air hole . In contrast, emission enhancement
at the red edge of the gap is only predicted for dipoles located
outside the air-hole, and preferentially for dipoles oriented
perpendicular to the air-hole edges. This difference between
dipole positions inside and outside the air holes is confirmed by
the 2D LRDOS modification predicted by the plane wave method
(Fig.~\ref{fig:2dpw}) and is consistent with the well-known
difference in spatial distribution of the low group-velocity modes
that cause the spontaneous emission enhancement. These modes are
preferentially concentrated in the air (dielectric) region for the
high (low) frequency edge of the gap. Disregarding the
Fabry-P\'erot oscillations due to the finite crystal  size, we
find that the spectral and spatial location of the enhancements,
the relative magnitude of the enhancements compared to the
  LRDOS modifications for frequencies far below the gap, as well as the band width of the gap in
Fig.~\ref{fig:contourLDOSmembrane}(a,b) are in good qualitative
agreement with the 2D LRDOS modification calculated by the
plane-wave method (Figure~\ref{fig:2dpw}).
It should be noted that good agreement is limited to dipole
orientations parallel to the membrane, as the vertical confinement
in the membrane causes a much larger shift to higher frequencies
for the  photonic band structure for TM polarization than for TE
polarization\cite{vuckovic}. The results in
Fig.~\ref{fig:contourLDOSmembrane} do not agree for any
polarization   with the much weaker 3D LRDOS modification for a
crystal that is infinitely extended in the vertical direction.

\section{Emission Gap width and the crystal size required for maximum emission gap depth\label{gapwidth}}
\begin{figure}[tb]
\centerline{\includegraphics[width=\figwidth]{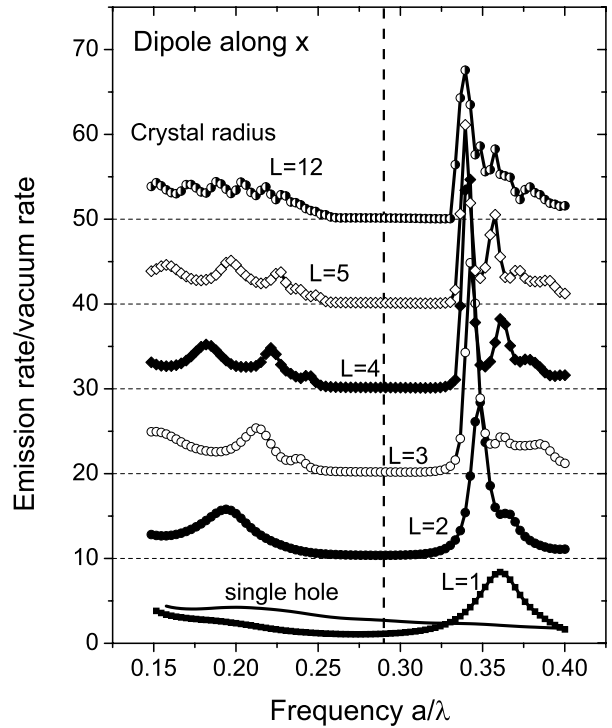}}\caption{Lines
with symbols: LRDOS normalized to vacuum LRDOS versus frequency,
depending on crystallite size (crystal  radii $L=1$ to 12 as
indicated, curves shifted vertically, and different symbols as
labelled) for an $x$-oriented dipole in the central air hole of
hexagonal crystallites (truncated with semiconductor). With
increasing number of layers, the gap deepens, the gap edges become
steeper and the spacing decreases for the Fabry-P\'erot fringes
that occur for frequencies below the gap $a/\lambda < 0.25$ and
above the gap $a/\lambda> 0.32$. For a single air hole in a
semiconductor (line, no symbols) no gap or resonance is observed,
while  for a  single crystal layer (small dots) a resonance at
frequencies above the gap is already evident. The vertical dashed
line marks the mid-gap frequency for which the inhibition is
reported in Fig.~\ref{fig:gapsize}. \label{fig:LDOSasafsize}}
\end{figure}

Figures~\ref{fig:centralinhib} and \ref{fig:contourLDOSmembrane}
allow us to conclude that large changes in spontaneous emission
rate occur  for properly oriented dipoles   located near the
center of small crystals in semiconductor membranes. In this
section we investigate  how large the crystal around the dipole
must be to obtain a significant emission gap. In
Fig.~\ref{fig:LDOSasafsize} the spontaneous emission modification
is shown for a horizontally $x$-oriented dipole in the central air
holes of   semiconductor-truncated crystals of various sizes. The
most striking features in Fig.~\ref{fig:LDOSasafsize} are the
deepening of the emission gap with increasing size of the
crystallite around the source, the steepening of the emission gap
band edges, and the decrease of the frequency spacing between
modulations of the emission rate outside the emission gap.
 The decreasing frequency
spacing between these fringes with increasing crystal size
validates their interpretation as Fabry-P\'erot type resonances
caused by the impedance mismatch occurring at the outer edges of
the crystal.   When crystals are truncated by air rather than
semiconductor, the oscillations change in amplitude and shift in
frequency at unchanged frequency   spacing between fringes, as
consistent with an interpretation as  Fabry-P\'erot resonances.
Unlike the Fabry-P\'erot oscillations that occur for transmission
through finite crystals\cite{shung93,norrisapl}, however, the
oscillations in the LRDOS depend on \emph{all\/} size-dependent
resonances in the system, instead of solely those with specific
wave vectors. Independently of the size and nature of the
truncation, the LRDOS
 at the high frequency edge of the emission gap exceeds
$15$ times the vacuum LRDOS, corresponding to  an enhancement of
at least   $5$ times compared to  the emission rate typically
observed below the gap in the range $0.15\leq a/\lambda\leq 0.22$.

\begin{figure}[tb]
\centerline{\includegraphics[width=\figwidth]{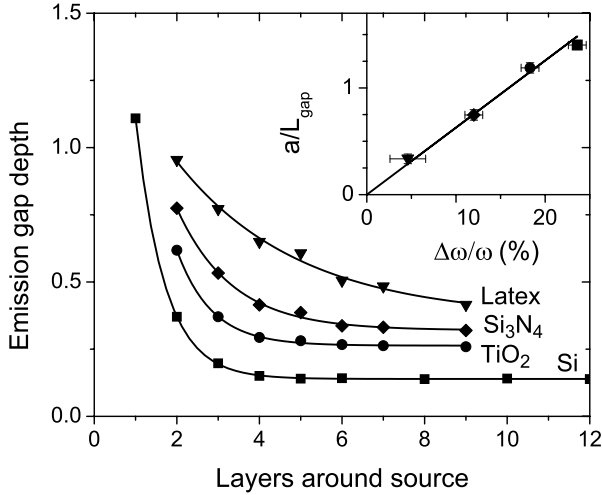}}\caption{Solid
squares: Spontaneous emission rate compared to vacuum for the
mid-gap frequency labelled in Fig.~\ref{fig:LDOSasafsize} versus
crystallite radius. Circles, diamonds and triangles: emission gap
depth in the middle of the emission gap for membranes (thickness
$250$~nm, $a=250$~nm, $r/a=0.3$) of TiO$_2$ ($\epsilon=6.25$),
Si$_3$N$_4$ ($\epsilon=4$), and latex ($\epsilon=2.25$). Lines are
exponential fits to the points. The inset shows the inverse fitted
exponential decay lengths (symbols, in units of $a$) versus the
relative bandwidth of the emission gap. The line corresponds to
$a/L_{\rm gap}=2\pi \Delta \omega/\omega$. \label{fig:gapsize}}
\end{figure}

 The
value of the LRDOS for frequencies within the emission gap
saturates with increasing crystal size. In Fig.~\ref{fig:gapsize},
we  plot the emission inhibition gap minimum for the frequency
$a/\lambda=0.29$ in the middle of the emission gap as a function
of crystal  radius. The LRDOS in the middle of the gap decreases
exponentially\cite{asatryan01,hermann,kole} to its limiting value
of 0.14~times the vacuum value, with a $1/e$ crystal radius of
just $L_{\rm gap}=0.7a$ . It is remarkable that only so few
crystal layers are needed to create the emission gap. This
observation suggests that the emission gap can be expected to be
very robust against fabrication disorder. The $1/e$ radius for the
emission gap can be compared to the number of layers  needed for a
cavity to reach its ultimate quality factor. The inverse quality
factor of low volume microcavities in 2D photonic crystals
approaches its limiting value  set by the out-of-plane losses in
an exponential manner as the number of crystal rows surrounding
the cavity is increased\cite{vuckovic,nodaprb}. For unoptimized
and optimized single-hole
cavities~\cite{srinivasan,nanoswitch,vuckovic} operating at
frequencies in the middle of the band gap, the same characteristic
length of $0.7a$ is found for the inverse quality factor.

 Figure~\ref{fig:gapsize} also shows the gap
minimum as a function of    crystal  size for photonic crystal
membranes  of different dielectric constants $\epsilon=6.25$
(titanium dioxide, $a=250$~nm), $\epsilon=4.0$ (silicon nitride,
$a=250$~nm) and $\epsilon=2.55$ (polystyrene, $a=250$~nm). As
expected, the maximum achievable inhibition decreases with reduced
index contrast, and the characteristic number of layers $L_{\rm
gap}$ needed to achieve the inhibition increases with decreasing
refractive index contrast. The inset in Fig.~\ref{fig:gapsize}
shows the inverse characteristic length $L_{\rm gap}$ as a
function of the relative band width $\Delta\omega/\omega$ of the
emission gap. We observe that $L_{\rm gap}$ is inversely
proportional to the width of the band
 gap. This inverse dependence is well known for the penetration depth of light in dielectric mirrors, and for
 simple two-band models of Bragg
 diffraction from photonic crystal lattice planes~\cite{shung93}. In such a model the field intensity decays exponentially into the
 crystal, with intensity decay length $D/(\pi \Delta
 \omega/\omega)$ where  $D$ is the separation between lattice
 planes. The same behavior is found for $L_{\rm gap}$, as shown by quantitative comparison of
 $L_{\rm gap}$ taken from the FDTD simulations with $a/(2\pi \Delta
 \omega/\omega)$.
  This observation highlights the universal role of the
 gap width as a measure for the interaction strength of
 the photonic structure with light, as proposed in
 Ref.~\onlinecite{kole,vos96}. A more
thorough investigation of this hypothesis should  include
calculations for 2D and 3D photonic crystal geometries and  is
outside the scope of this paper.

\section{Different contributions to the emission rate  derived  from a simple homogeneous slab model\label{sepdep}}
\begin{figure}[tb]
\centerline{\includegraphics[width=\figwidth]{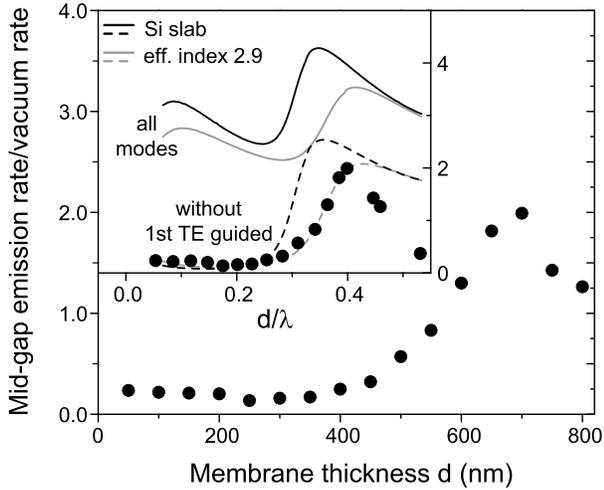}}
\caption{Solid circles: emission rate normalized to the vacuum
rate at the center frequency of the emission gap for an
$x$-oriented dipole in the center of a photonic crystal membrane
as a function of membrane thickness $d$. For membranes above
$500$~nm thickness (around one third of the emission wavelength)
no inhibition of emission remains. Inset: (symbols) the same data
plotted as a function of thickness normalized to the gap center
wavelength (which shifts with thickness). Solid lines: LRDOS for
an in-plane oriented dipole in the center of a homogeneous
membrane of refractive index 3.45 (black) and of index $2.9$
(gray). Dashed lines: same as solid lines but without the
contribution of the first TE-polarized slab guided mode that is
suppressed by the 2D photonic band gap.
\label{fig:depththickness}}
\end{figure}

Figure~\ref{fig:gapsize} not only shows that the emission gap in
membrane photonic crystals deepens exponentially with  lateral
crystal size, but also  that the inhibition that is ultimately
achieved for sufficiently large crystals depends on the membrane
index. Similarly, we expect the emission gap depth to depend on
the thickness of the membrane. We have calculated the spontaneous
emission rate for a dipole in the central air hole of photonic
crystal membranes as considered in Fig.~\ref{fig:centralinhib}
($a=420$~nm, $r/a=0.3, \epsilon=11.76,L=6$) for membrane
thicknesses between $50$ and $800$~nm.
Figure~\ref{fig:depththickness}  shows the emission rate
normalized to the vacuum rate at the center frequency of the
emission gap,  which shifts from $a/\lambda=0.45$ at $d=50$~nm to
lower frequencies as the membrane  thickness increases. The center
frequency of the gap saturates at around $a/\lambda=0.26$  for
thicknesses around $400$~nm. For membrane thicknesses above
$600$~nm contributions of higher order slab guided modes
compromise the definition of the gap edges that allows us to find
the center frequency. For these thicknesses  the Fabry-P\'erot
oscillations for frequencies just below the red edge of the gap
merge with Fabry-P\'erot oscillations for in-gap frequencies. The
enhancement at the blue edge of the gap remains clearly visible
and remains above 10 times the vacuum LRDOS.  For thicknesses
above $700$~nm the emission rate  modulation in the frequency
range around the gap  approaches an asymmetric  sawtooth like
shape, similar to the 3D LRDOS in a vertically infinite crystal
(Figure.~\ref{fig:centralinhib}).

Figure~\ref{fig:depththickness} shows that significant inhibition
of emission occurs for membrane thicknesses below $400$~nm, with
an optimum thickness for inhibition around $250$~nm. For larger
thicknesses, the emission rate is significantly less inhibited,
and even enhanced for $d\geq 600$~nm. As a simple model to explain
the thickness dependence of the emission rate, we consider the
emission rate of dipoles in homogeneous dielectric slabs. We have
used the approach of Urbach and Rikken~\cite{urbach} that allows
to separate the contributions to the LRDOS of modes that propagate
outside the   dielectric slab from the contributions of the slab
guided modes to the lifetime of in-plane oriented dipoles. In a
high index dielectric slab, such as a silicon membrane in air
(Fig.~\ref{fig:depththickness}, inset) the LRDOS contribution of
propagating modes is strongly reduced   to only $\sim 0.08$ times
the vacuum LRDOS. However, the LRDOS contribution of the guided
modes causes the overall LRDOS, and hence the spontaneous emission
rate for an embedded dipole to exceed the vacuum LRDOS
significantly.

For a photonic crystal membrane  that has   a full band gap for
the first TE slab-guided mode, one expects  the LRDOS contribution
of the first TE guided mode to vanish completely.  To assess the
validity of this hypothesis, we plot   the mid-gap emission rate
in Fig.~\ref{fig:depththickness}  as a function of normalized
thickness  $d/\lambda$   in the inset of
Fig.~\ref{fig:depththickness} and compare it to the LRDOS in a
homogeneous silicon slab, \emph{without} the contribution of the
first TE guided mode. The comparison reveals that the emission
rate in the emission gap of Si photonic crystal membranes  indeed
follows qualitatively the same dependence on slab thickness as the
LRDOS in a homogeneous silicon membrane  {without} the
contribution of the first TE guided mode, yet shifted towards
larger thicknesses. The sharp increase of the emission rate with
thickness around $d/\lambda=0.3$  allows us to fit an effective
refractive index of $n_{\rm eff}\sim 2.9$ for which the LRDOS at
gap frequencies of the membranes most resembles the LRDOS without
the first TE guided mode contribution  in   unpatterned dielectric
slabs. This value is  higher than the effective index of $\sim
2.8$ derived from the volume averaged dielectric constant of the
membrane. It is important to note, however, that the  sharp
increase of the emission rate is due  to the combined
contributions of the second order TM and the third order TE  slab
guided mode, for which the derivation of an effective index is
highly nontrivial.

\begin{figure}[tb]
\centerline{\includegraphics[width=\figwidth]{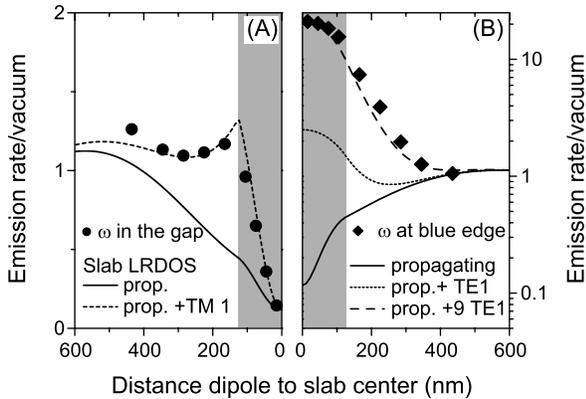}}
\caption{(A): Symbols: Emission rate normalized to vacuum emission
rate as a function of the vertical separation between dipole and
the center of the membrane for an $x$-oriented dipole in  the
central hole   of the structure considered in
Fig.~\ref{fig:centralinhib}, emitting at a frequency
$a/\lambda=0.28$ inside the emission gap.  At the inside edge of
the membrane and outside the membrane, no trace of inhibition
remains. Solid line: LRDOS contribution of propagating modes for
an in-plane oriented dipole in the center of a homogeneous
dielectric slab (index $2.8$) with the same thickness as the
membrane. Dashes: same as solid line, augmented with the
contribution of the first TM guided mode to the LRDOS. (B)
Symbols: as in (A) but for a frequency $a/\lambda=0.33$ at the
blue edge of the gap. Solid line: propagating mode  LRDOS
contribution as in (A). Dotted line: same as solid line, augmented
with the contribution of the first TE guided mode to the LRDOS of
a homogeneous slab. Dashed line: same as solid line, augmented
with a nine times enhanced contribution of the first TE guided
mode. The extent of the membrane is shaded in gray in (A) and (B).
 \label{fig:sepdep}}
\end{figure}

The interpretation of lifetimes in photonic crystal membranes in
terms of the propagating and guided mode LRDOS contributions in a
dielectric slab  is also relevant for the dependence of the LRDOS
on the separation of the dipole from the center of the photonic
crystal membrane. In Fig.~\ref{fig:sepdep} we plot the emission
rate at two key frequencies for an $x$-oriented dipole  in the
central air hole of a  membrane of $250$~nm thickness
  that is shifted vertically   above
the center of the slab. For a frequency $a/\lambda=0.28$ in the
emission gap (Fig.~\ref{fig:sepdep}(a)) we find that the
inhibition of emission quickly vanishes as the dipole is moved
away from the center of the  slab. No traces of inhibition remain
for dipoles that are further than $\sim 100$~nm away from the slab
center. In contrast, we find that the enhancement at
$a/\lambda=0.33$ at the blue edge of the gap is not restricted to
dipole positions within 100~nm of the slab center. Emission
enhancement by a factor $5$ persists even  $50$~nm above the slab,
as we proposed recently~\cite{shortldos}

The  differences in    persistence of the inhibition and of the
enhancement of spontaneous emission with increasing separation of
the emitter to the  slab can   also be understood by analyzing
the various guided mode contributions to the LRDOS  of a
homogeneous slab. For the frequency $a/\lambda$ in the gap, no
contribution to the LRDOS of the first TE guided mode exists, due
to the TE band gap. In this frequency range, the only other modes
that contribute to the LRDOS in a dielectric slab are the
propagating modes and the first TM guided mode. The electric field
component  parallel to the membrane of the first TM guided mode is
zero in the center of the slab, but increases monotonically
towards the edge of the slab. Indeed, we find excellent agreement
of the LRDOS calculated by FDTD  for $x$-oriented  dipoles in the
membrane, and the sum of LRDOS contributions of propagating modes
and the first TM guided mode in a dielectric slab of index $2.8$,
equal to the volume average refractive index of the photonic
crystal. In essence, the range of dipole heights over which
inhibition occurs is limited by the parallel electric field
profile of the first TM slab mode. In contrast, the enhancement of
emission at the blue edge of the gap is due to the photonic
crystal induced enhanced density of first order TE guided modes.
The emission enhancement as a function of dipole height therefore
traces the vertical mode profile of the first TE guided mode. This
notion is confirmed by the fact that the enhancement in
Figure~\ref{fig:sepdep}(b)  for a PC membrane, can be excellently
described as the sum of the  propagating mode LRDOS contribution
in the effective index slab, augmented by a scaled contribution of
the first TE guided mode. Quantitatively, we find that the
contribution of the first TE guided mode is enhanced by $\sim 9$
times over the contribution of the same mode in a homogeneous slab
of the same effective index.

\section{Orientational dipole ensembles\label{ensemble}}
Single quantum emitters that are controlled in position and dipole
orientation are required to experimentally confirm the strong
dependence on position and polarization of the spontaneous
emission rate predicted by the simulations. However, the
preparation, manipulation and detection of high-quantum efficiency
single emitters face many experimental challenges in the
near-infrared range where most photonic crystals operate best. In
such experiments, we therefore expect that  ensembles of emitters
are likely to be used\cite{olkapl}.   In this section we evaluate
the emission rate modification that is observable  for a
subwavelength-sized orientationally isotropic ensemble of dipoles
located in the central air hole of the photonic crystal membrane
considered in Figs.~\ref{fig:centralinhib} and
\ref{fig:contourLDOSmembrane}. To obtain an angle-averaged
modification of the emission rate for an isotropic ensemble of
dipole orientations, one often uses a simple average of the LRDOS
over three perpendicular dipole orientations\cite{urbach}.
However, because of the angle-dependence of the  luminescence
extraction efficiency\cite{fanextract} $\eta$, such a simple
average is not suited for reconstructing the ensemble-averaged
emission rate modification that is observable with far-field
collection optics. The fraction of light emitted into propagating
waves instead of the guided mode of the slab  depends strongly on
dipole orientation. For example, a dramatic increase of the
emission extraction efficiency  from $\sim 20\%$ for frequencies
below the gap to above 80\% in the gap occurs for in-plane
dipoles. This increase counteracts the reduction in visibility of
the emission gap that is naively expected due to the contribution
of vertically oriented dipoles.

\begin{figure}[tbh]
\centerline{\includegraphics[width=\figwidth]{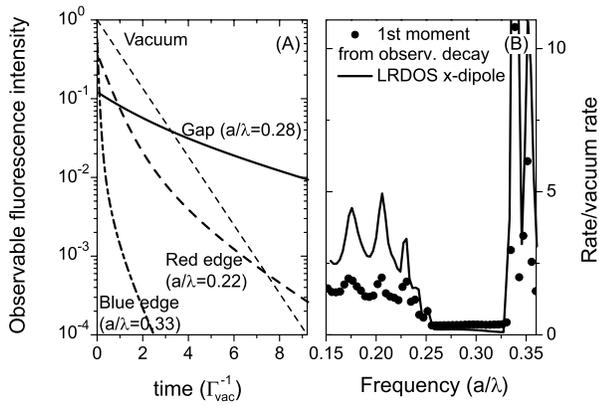}}
\caption{\label{fig:ensemble}(A) Simulated  fluorescence decay
curves (taking extraction efficiency into account) versus time (in
units of the vacuum decay time $\Gamma_{\rm vac}^{-1}$) for an
isotropically oriented ensemble of dipoles in the center of the
photonic crystal membrane crystal  for three frequencies (below
the gap, $a/\lambda=0.23$ (dash-dotted), in the gap
$a/\lambda=0.28$ (solid), above the gap $a/\lambda=0.33$
(dashed)). The dotted line shows the decay of the same ensemble in
vacuum. (B) Dots: first moment of simulated decay curves versus
emission frequency, in units of the vacuum decay rate. Line:
spontaneous emission rate over vacuum rate for a single dipole in
the center of the membrane reproduced from
Fig.~\ref{fig:centralinhib}. }
\end{figure}

 To assess the changes in the emission dynamics observable in
time-resolved fluorescence measurements, we have calculated the
LRDOS and $\eta$ for  differently oriented dipoles in the central
air hole. We obtained  $\eta$ from the frequency-dependent
Poynting vector using on-the-fly discrete Fourier transform of the
time-dependent tangential E and H-fields on the surface of a box
enclosing the crystal structure.\cite{taflove} To synthesize the
observable ensemble-averaged decay dynamics, one needs to average
exponential time traces with decay rate set by the LRDOS and
weighting factor proportional to $\eta$ over all possible dipole
orientations (described by polar angles $\phi$ and $\theta$):
$$
\langle I(t)\rangle =\frac{I_0}{2\pi}
\int_{0}^{2\pi}\int_0^{\pi/2}
\eta(\theta,\phi){\gamma(\theta,\phi)} {e^{-\gamma(\theta,\phi)
t}} \sin(\theta) d\theta d\phi.
$$
Here, time $t$ is in units of the vacuum decay time, and
$\gamma(\theta,\phi)$ is the orientation-dependent emission rate
normalized to the vacuum emission rate. We have approximated this
integral by averaging over more than 300 orientations in $2\pi$
solid angle, corresponding to over 25 inequivalent dipole
orientations. In general, we obtain strongly non-single
exponential time traces for the ensemble decay.
Fig.~\ref{fig:ensemble}(a) presents the fluorescence decay for
three key frequencies $a/\lambda=0.22, 0.26$ and $0.33$  just
below, in, and just above the emission gap. At the red edge of the
gap the non-exponential decay reveals both an enhanced and an
inhibited decay component. For frequencies inside the gap, we
predict that the ensemble-averaged decay exhibits a clear slowing
of the decay dynamics, while a substantial acceleration is
predicted to be observable for frequencies at the blue edge of the
gap. The non-exponential nature of the time traces, which for some
frequencies is   more pronounced than for the three traces shown
in Fig.~\ref{fig:ensemble}(a), requires  that experiments  be
performed with sources for which the decay dynamics in a medium of
unmodulated LRDOS is completely understood. Furthermore  an
analysis method to quantify the overall emission rate modification
will be necessary. Here, we consider   the first moment
$\bar{\tau}$
$$ \bar{\tau}=\frac{\int_0^\infty t\langle I(t)\rangle
dt}{\int_0^\infty \langle I(t)\rangle dt}
$$
of the synthesized fluorescence decay to quantify the overall
emission rate modification evident in the nonexponential time
traces. In Fig.~\ref{fig:ensemble}(b) we plot the inverse
$\bar{\tau}^{-1}$ of the first moment in units of the vacuum decay
rate for different emission frequencies. We see that even the
dipole orientation-averaged  emission will reveal inhibition
(enhancement) of the mean lifetime by a factor 3 (5) compared to
vacuum.

\section{Conclusions\label{conclusions}}
We have used the three-dimensional Finite Difference Time Domain
method to systematically investigate the spontaneous emission rate
modifications that occur for emitters inside photonic crystal
membranes.  The  spontaneous emission rate  depends strongly on
dipole position and orientation relative to the crystal lattice.
In the central plane of the slab and for in-plane oriented
dipoles, the full gap for the first TE  guided mode of the
vertical structure  causes an inhibition of emission by more than
seven  times compared to vacuum over a large frequency window. In
addition,  large  enhancements of the emission rate occur at the
band edges. In the middle of the membrane, vertical index
confinement causes the emission rate modifications  to be much
stronger than in vertically unconfined 2D crystals.  The true 3D
LRDOS in a photonic crystal membrane mimics the hypothetical 2D
LRDOS of a 2D crystal in terms of the frequency band width of the
gap, the magnitude of the enhancement (with respect to vacuum) at
the band edges, and the variations of the LRDOS spectrum with
lateral position of the dipole in the unit cell. While the LRDOS
in the central plane of the membrane can be understood from the 2D
band structure, the variation of the LRDOS with the height of the
dipole from the center of the slab can be explained by considering
the electric field mode profiles of the lowest few TE and TM
guided modes supported by the thin membrane. This is especially
relevant for applications that aim at enhancing spontaneous
emission. Due to the vertical extent of the first TE guided mode
of the membrane, sizeable emission rate enhancements due to the
photonic crystal remain even outside the membrane. For various PC
structures of different index contrasts, we have found that the
emission rate inside the emission gap reduces exponentially with
the system radius to a residual value that can be understood as
the LRDOS contribution of modes propagating in air for a
homogeneous dielectric slab of proper effective index. We have
furthermore presented evidence that the $1/e$ crystal radius
associated with the deepening of the emission gap scales inversely
with the relative frequency band width of the gap in the 2D
dispersion relation.  Finally, we have also considered the role of
ensemble averaging in experiments that use localized collections
of emitters. A realization of such an experiment involves, for
instance, a sub-wavelength ensemble of emitters\cite{olkapl}
attached to the end of a sharp glass tip and positioned by using
scanning probe techniques, acting  as a controllable local probe
of the emission modification. Such a sub-wavelength source
averages over an isotropic random distribution of dipole
orientations. Despite the presence of dipole orientations that do
not match the gap for TE modes, we predict   the first moment of
the expected non-exponential times traces to show sizeable
inhibition and enhancement of the mean emission rate.

\section*{Acknowledgments} We are grateful to Ben Buchler for
stimulating discussion. Financial support by EU-projects DALHM,
Metamorphose, Phoremost, as well as by
 DARPA (Contract number  MDA972-01-2-0016)and Ames Laboratory
are also acknowledged. This project was performed within the focus
program SP1113 of the Deutsche Forschungsgemeinschaft (DFG).
Femius Koenderink's current address is FOM Institute for Atomic
and Molecular Physics, Kruislaan 407 1098 SJ, Amsterdam, The
Netherlands. He can be reached by e-mail at F.Koenderink@amolf.nl,
and Vahid Sandoghdar can be reached by e-mail at
vahid.sandoghdar@ethz.ch.


\begin{thebibliography}{99}


\bibitem{drexhage70}{K. H. Drexhage, ``Influence of a Dielectric Interface on Fluorescence Decay Time'',
   {J. Lumin.} \textbf{1}, 693--701 (1970).}

\bibitem{khosravi91}{H. Khosravi and R. Loudon,
``Vacuum Field Fluctuations and Spontaneous Emission in the
Vicinity of a Dielectric Surface,''
  Proc. Roy. Soc. London  \textbf{433},
  337-- 352 (1991)
}



\bibitem{snoeks95}{E. Snoeks, A. Lagendijk, and A. Polman, ``Measuring and modifying radiative transitions rate of erbium near
an interface,'' Phys. Rev. Lett. \textbf{74}, 2459-2462 (1995).}

\bibitem{amosandbarnes}R. M. Amos and W. L. Barnes, ``Modification of the spontaneous emission rate of Eu$^{3+}$
 ions close to a thin metal mirror,'' Phys. Rev. B  \textbf{55}, 7249--7254
 (1997).


\bibitem{buchler05}{B. C. Buchler, T. Kalkbrenner, C. Hettich, and
V. Sandoghdar, ``Measuring the quantum efficiency of single
radiating dipoles,'' Phys. Rev. Lett. \textbf{95}, 063003:1--4
(2005). }


\bibitem{qbook} P. R. Berman, ed., {\em Cavity Quantum Electrodynamics} (Academic,
San Diego, Calif., 1994).

\bibitem{chew} H. Chew, ``Radiation and lifetimes of atoms inside dielectric particles,'' Phys. Rev. A \textbf{38}, 3410--3416,
(1988).

\bibitem{klimov} V. V. Klimov, M. Ducloy, and V. S. Letokhov, ``Spontaneous emission rate and level shift of an atom inside a dielectric
microsphere,'' J. Mod. Opt. \textbf{43} 549--563 (1996).

\bibitem{hermannssphereref} F. L. Kien, N. H. Quang, and K. Hakuta,
``Spontaneous emission from an atom inside a dielectric sphere,''
 Opt. Commun. \textbf{178}, 151--164 (2000).


\bibitem{schniepp} H. Schniepp, and V. Sandoghdar, ``Spontaneous Emission of Europium Ions Embedded in Dielectric
Nanospheres,''
 Phys. Rev. Lett. \textbf{89}, 257403:1-4 (2002).




\bibitem{vahalareview}{K. J. Vahala, ``Optical Microcavities'',
Nature \textbf{424}, 839--846 (2003).}



\bibitem{soukoulis01}
C.~M. Soukoulis, ed., \emph{Photonic crystals and light
localization in the
  $21^{\rm st}$ century} (Kluwer Academic Publishers, Dordrecht, 2001).


\bibitem{bykov72}{V. P. Bykov,
``Spontaneous Emission in a Periodic Structure,''
  Sov. Phys.--JETP
\textbf{35}, {269--273} (1972). }


\bibitem{yablonovitch87}{E. Yablonovitch,``{Inhibited Spontaneous Emission in Solid-State Physics and
          Electronics},''
Phys. Rev. Lett. \textbf{58}, 2059--2062 (1987). }


\bibitem{blackbody}C. M. Cornelius and J. P. Dowling, ``Modification of Planck blackbody radiation by photonic band-gap
structures'', Phys. Rev. A \textbf{59}, 4736--4746 (1999).





\bibitem{slowing} M. F. Yanik and S. Fan, ``Stopping light all optically,'' Phys. Rev. Lett. \textbf{92}, 083901:1-4 (2005).


\bibitem{koenderink02}
A.~F. Koenderink, L.~Bechger, H.~P. Schriemer, A.~Lagendijk, and
W.~L. Vos, ``{Broadband fivefold reduction of vacuum fluctuations
probed by dyes
  in photonic crystals,}'' Phys. Rev. Lett. \textbf{88}, 143903:1-4 (2002).


\bibitem{lodahl04}{
P.~Lodahl, A.~F. van Driel, I.~S. Nikolaev, A.~Irman, K.~Overgaag,
D.~Vanmaekelbergh, and W.~L. Vos, ``{Controlling the dynamics of
spontaneous emission from
 quantum dots by photonic crystals,}'' Nature \textbf{430}, 654--657
 (2004).}


\bibitem{ogawa04} S. P. Ogawa, M. Imada, S. Yoshimoto, M. Okano, S.
Noda, ``Control of light emission by 3D photonic crystals,''
Science \textbf{305}, 227--229 (2004).

\bibitem{nodacav}{Y. Akahane, T. Asano, B. S. Song, S. Noda, `` High-Q photonic nanocavity in a two-dimensional photonic crystal,''  Nature \textbf{425},  944--947 (2003). }


\bibitem{srinivasan} K. Srinivasan, P. E. Barclay, O. Painter, J. X. Chen, A. Y.
Cho, C. Gmachl, ``{Experimental demonstration of a high quality
factor photonic crystal microcavity},'' Appl. Phys. Lett.
\textbf{83}, 1915--1917 (2003).

\bibitem{yoshie}{T. Yoshie, A. Scherer, J. Hendickson, G. Khitrova, H.M. Gibbs, G. Rupper, C. Ell, O.B. Shchekin D.G. Deppe,
``{Vacuum Rabi splitting with a single quantum dot in a photonic
crystal nanocavity},'' Nature \textbf{432}, 200--203 (2004).}


\bibitem{finley}{A. Kress, F. Hofbauer, N. Reinelt, M. Kaniber, H. J. Krenner, R. Meyer, G. Bohm, and J. J. Finley,
``Manipulation of the spontaneous emission dynamics of quantum dots in two-dimensional photonic crystals,''
 Phys. Rev. B \textbf{71} 241304, 2005.}


\bibitem{vuckovicemits}{
 D. Englund, D. Fattal, E. Waks, G. Solomon, B. Zhang, T. Nakaoka, Y. Arakawa, Y. Yamamoto, J.
 Vu\v{c}kovi\'c
``Controlling the Spontaneous Emission Rate of Single Quantum Dots
in a 2D Photonic Crystal,''  Phys. Rev. Lett. \textbf{95} 013904 (2005).}

\bibitem{nodainhibit}{M. Fujita, S. Takahashi, Y. Tanaka, T. Asano, S. Noda ,
``Simultaneous inhibition and redistribution of spontaneous light
emission in photonic crystals'', Science \textbf{308},
1296-1298 (2005).}



\bibitem{ho90}
K. M. Ho, C. T. Chan, and C. M. Soukoulis, ``Existence of a
photonic gap in periodic dielectric structures,'' Phys. Rev. Lett.
\textbf{65}, 3152--3155 (1990).

\bibitem{suzuki95}{T. Suzuki and P. K. L. Yu, ``Emission power of an electric
dipole in the photonic band structure of the fcc lattice,'' J.
Opt. Soc. Am. B \textbf{12}, 570--582 (1995).}



\bibitem{busch98}{
K.~Busch and S.~John, ``{Photonic band gap formation in certain
  self-organizing systems},'' Phys. Rev. E \textbf{58}, 3896--3908 (1998).
}


\bibitem{busch00}{K. Busch, N.  Vats, S. John and  B. C.
Sanders, ``{Radiating Dipoles in Photonic Crystals},'' {Phys. Rev.
E}  \textbf{62}, 4251--4260 (2000).}



\bibitem{li00} Z. Y. Li, L. L. Lin, and Z. Q. Zhang, ``Spontaneous emission
from photonic crystals: full vectorial calculations,'' Phys. Rev.
Lett. \textbf{84}, 4341--4344 (2000).



\bibitem{wanggrouptheory}R. Z. Wang, X. H Wang, B. Y Gu, and G. Z.
Yang, ``{Local density of states in three-dimensional photonic
crystals: calculation and enhancement effects},'' Phys. Rev. B
\textbf{67}, 155114:1-7 (2003).



\bibitem{hwang99} J.-K. Hwang, H.-Y. Ryu, and Y.-H. Lee, ``Spontaneous emission
rate of an electric dipole in a general microcavity,'' Phys. Rev.
B \textbf{60}, 4688--4695 (1999).


\bibitem{xu00}{Y. Xu, R. K. Lee,  and A. Yariv,
``Quantum Analysis and the Classical Analysis of Spontaneous
Emission in a Microcavity,'' Phys. Rev. A \textbf{61},  033807:1-13
(2000).}

\bibitem{lee00}{R. K. Lee, Y. Xu, A. Yariv, ``Modified Spontaneous Emission from a two-dimensional Photonic Bandgap Crystal Slab,'' J. Opt. Soc. Am. B \textbf{17}, 1438--1442 (2000).}

\bibitem{hermann}{C. Hermann and O. Hess, ``Modified Spontaneous Emission rate in an Inverted-opal Structure with Complete Photonic Bandgap,'' J. Opt. Soc. Am. B \textbf{19}, 3013--3018 (2002).}


\bibitem{taflove}{A. Taflove and S.C. Hagness, \emph{Computational
Electrodynamics: The Finite-Difference Time-Domain Method}  (2nd
ed., Artech House, Boston, MA, 2000).}




\bibitem{sprik96}{R. Sprik, B. A.  van Tiggelen,  and A.
Lagendijk, ``Optical Emission in Periodic Dielectrics,'' Europhys.
Lett.  \textbf{35},
   265--270 (1996).}






\bibitem{daubechies}{I. Daubechies, ``Orthonormal Bases of Compactly Supported Wavelets,''
Comm. Pure Appl. Math. \textbf{41}, 909-996 (1988).}

\bibitem{nanoswitch}{A. F. Koenderink, M. Kafesaki, B. C. Buchler, V.
Sandoghdar, ``Controlling the Resonance of a Photonic Crystal
Microcavity by a Near-Field Probe'', Phys. Rev. Lett. \textbf{95},
153904:1--4 (2005).}



\bibitem{hermannpssa}{O. Hess, C. Hermann and A. Klaedtke, ``Finite-Difference Time-Domain simulations of photonic crystal defect structures,''
Phys. Stat. Sol. A \textbf{197}, 605--619 (2003).}






\bibitem{gilat72}{G. Gilat,``Analysis of Methods for Calculating Spectral Properties in Solids,''
  J. Comput. Phys. \textbf{10},
 {432--465} (1972).
}

\bibitem{monkhorst76}{H. J. Monkhorst, and J. D. Pack},
  ``{Special Points for Brillouin-Zone Integrations},''
Phys. Rev. B  \textbf{13}, 5188--5192 (1976).





\bibitem{asatryan01}  A. A. Asatryan, K. Busch, R. C. McPhedran, L. C.  Botten,
           C. M. de Sterke, . and N. A. Nicorovici,
``{Two-Dimensional {G}reen's Function and Local Density of States
in Photonic Crystals Consisting of a Finite Number of Cylinders of
Infinite Length},''
 {Phys. Rev. E} \textbf{63}, 046612:1-4 (2001).






\bibitem{fussell03}{D. P. Fussell, R. C. McPhedran, C. M.  de Sterke, and A. A.
Asatryan, ``Three-dimensional local density of states in a finite
two-dimensional photonic crystal composed of cylinders,'' Phys.
Rev.  E \textbf{67}, 045601:1-4 (2003). }

\bibitem{vuckovic} O. Painter, J.
Vu\v{c}kovi\'c, and A. Scherer, ``Defect modes of a
two-dimensional photonic crystal in an optically thin dielectric
slab,'' J. Opt. Soc. Am. B \textbf{16}, 275--285 (1999).



\bibitem{shung93}{K.W.-K. Shung and Y. C. Tsai, ``Surface effects and band
measurements in photonic crystals,''  Phys. Rev. B \textbf{48},
11265-11269 (1993).}



\bibitem{norrisapl}{ Yu. A. Vlasov, M. Deutsch, and D. J. Norris, ``Single-domain spectroscopy of self-assembled photonic
crystals,'' Appl. Phys. Lett. \textbf{76}, 1627--1629 (2000).}




\bibitem{kole}{ J. S. Kole, \emph{New methods for the numerical solution of Maxwell's
equations,} (PhD thesis, Rijksuniversiteit Groningen, The
Netherlands (2003))
http://www.ub.rug.nl/eldoc/dis/science/j.s.kole/}



\bibitem{nodaprb}{ M. Okano, A. Chutinan, and S. Noda, ``Analysis and design of single-defect cavities in a three-dimensional photonic
crystal,'' Phys. Rev. B \textbf{66}, 165211:1-6 (2002).}




\bibitem{vos96} W.L. Vos, R. Sprik, A. van Blaaderen, A. Imhof, A. Lagendijk, and
G.H. Wegdam, ``Strong effects of photonic band structures on the
diffraction of colloidal crystals,'' Phys. Rev. B. \textbf{53},
16231-16235 (1996).




\bibitem{urbach}{H. P. Urbach, and G. L. J. A. Rikken, ``Spontaneous emission from a dielectric slab,''
Phys. Rev. A \textbf{57}, 3913--3930 (1998).}



\bibitem{shortldos}{A. F. Koenderink, M. Kafesaki, C. M.
Soukoulis, V. Sandoghdar, ``Spontaneous emission in the near-field
of 2D  membrane photonic crystals,'' Opt. Lett, \textbf{30}, 3210--3212
(2005).}



\bibitem{olkapl} P. Olk, B. C. Buchler, V. Sandoghdar, N. Gaponik,
A. Eychm\"uller, A. L. Rogach, ``Subwavelength emitters in the
near-infrared based on mercury telluride nanocrystals,''
 Appl. Phys. Lett. \textbf{84},
 4732--4734 (2004).


\bibitem{fanextract}{S. Fan, P. R. Vileneuve, J. D. Joannopoulos, E. F. Schubert, %
``High extraction efficiency of epontaneous emission from slabs of photonic crystals'',
Phys. Rev. Lett. \textbf{78}, 3294--3297  (1997).}





\end{thebibliography}
\end{document}